\documentclass[reprint,aps,prb,amsmath,amssymb]{revtex4-1}

\usepackage[]{graphicx}
\usepackage{times}
\usepackage{bm}
\usepackage{color}
\renewcommand{\bm }{\mathbf}

\newcommand{\comment}[1]{}

\begin{document}

\title{Analytical approach to excitonic properties of $\mathbf{MoS_2}$}

\author{Gunnar Bergh\"auser}
\email[]{g.berghaeuer@tu-berlin.de}
\author{Ermin Malic}

\affiliation{Institut f\"ur Theoretische Physik, Nichtlineare Optik und Quantenelektronik, Technische Universit\"at Berlin, Hardenbergstr. 36, 10623 Berlin, Germany}

\begin{abstract}
We present an analytical investigation of the optical absorption spectrum of monolayer molybdenum-disulfide. Based on the density matrix formalism, our approach gives insights into the microscopic origin of excitonic 
transitions, their relative oscillator strength, and binding energy. 
We show analytical expressions for the carrier-light coupling element, which contains the optical selection rules and well describes the valley-selective polarization in $\text{MoS}_\text{2}$.
In agreement with experimental results, we find the formation of strongly bound electron-hole pairs due to the efficient Coulomb interaction. The absorption spectrum of $\text{MoS}_\text{2}$ on a silicon substrate features two pronounced peaks at 1.91 eV and 2.05 eV
corresponding to the A and B exciton, which are characterized by binding energies of 420 meV and 440 meV, respectively. Our calculations reveal their relative oscillator strength and predict the appearance of further low-intensity excitonic transitions at higher energies. The presented approach is applicable to other transition metal dichalcogenides and can be extended to investigations 
of trion and biexcitonic effects.
\end{abstract}

\maketitle

Transition metal dichalcogenides (TMDs) build a new class of layered two-dimensional materials
with remarkable optical and electronic properties.\cite{Butler2013} They show a crossover from indirect- to
direct-gap semiconductors depending on the thickness of the material.\cite{RadisavljevicBNatNano2011,HeinzPRL2010,Changgu2010,Splendiani2010}
Furthermore, they are characterized by a strong spin-orbit coupling that in combination with the circular dichroism \cite{WangPRB2008,HaiZhouPRLett2013} enables selective valley and spin polarization.\cite{CaoTing2012,HeinzNatNano2012,MakKinFai2013,Hualing2012,Sanfeng2013,SallenPRB2012} This makes TMDs
interesting for both fundamental research and technological applications.\cite{Butler2013}
In particular, monolayer molybdenum disulfide ($\text{MoS}_\text{2}$) has been intensively studied in the last years. It consists of Mo atoms sandwiched between two layers of S atoms, cf. Fig. 1 (c). Viewed from above, the structure builds a hexagonal lattice with alternating covalently bonded molybdenum and sulfur atoms, cf. Fig. 1 (b). Similar to graphene, MoS$_2$ shows strong many-body interactions resulting in a variety of physical phenomena.\cite{SaptarshiNanoLet2013,Yanyuan2013,RubioPRB2012,Hongyan2013} \\

In contrast to the bulk MoS$_2$, the monolayer material 
exhibits a direct gap giving rise to a strong photoluminescence, which is characterized by tightly bound excitons and even trion features have been observed.\cite{MakKinFai2013} So far, the experimental data 
\cite{Korn2011,Korn2012,Bratschitsch13,JinPRL2013, Yanyuan2013,MakKinFai2013,SallenPRB2012} has been complemented by a few calculations that significantly vary in their predictions with respect to the excitonic effects
.\cite{RamasubramaniamPRB2012,CheiwchanchamnangijPRB2012, SinghEPJB2012, SanchezPRB2013,BerkelbachPRB2013,Sun2013} 
Exploiting the Bethe-Salpeter equation combined with the G$_0$W$_0$ 
approximation, A. Ramasubramaniam et al.\cite{RamasubramaniamPRB2012} predicted the appearance of strongly bound excitons with a binding energy in the range of 1 eV. This is in agreement with the estimation by T. Cheiwchanchamnangij et al.\cite{CheiwchanchamnangijPRB2012} that relies on the Mott-Wannier effective-mass theory. In contrast, A. Molina-Sanchez et al.\cite{SanchezPRB2013} provided well converged optical spectra in the framework of
 Bethe-Salpeter including the spin-orbit coupling. They found a clearly weaker excitonic binding energy in the range of few hundreds of meV.

In this article, we present an analytical solution to the excitonic absorption spectrum of $\text{MoS}_\text{2}$. Based on the density matrix formalism, we derive the Wannier equation providing access to eigenvalues including higher excitonic transitions as well as excitonic eigenfunctions shedding light on the relative intensities of the single transitions. Our goal is a thorough understanding of the excitonic absorption spectrum of $\text{MoS}_\text{2}$ and related structures. 
\\

 \begin{figure}[t!]
 \begin{center}
\includegraphics[width=0.95\linewidth]{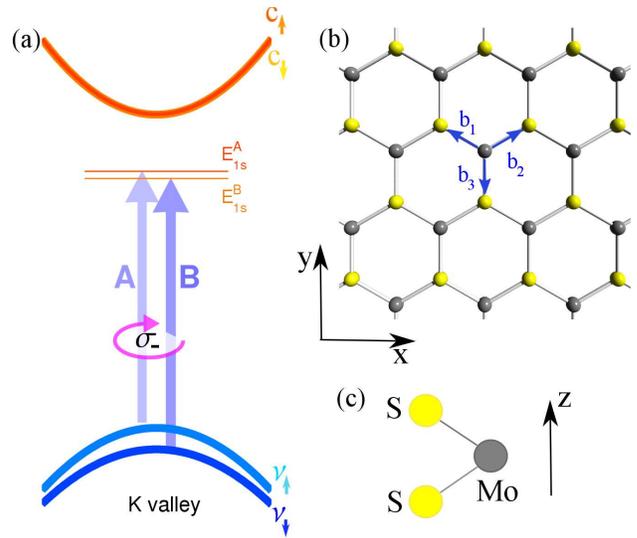}
 \end{center}
 \caption{(a) Bandstructure of $\text{MoS}_\text{2}$ in the vicinity of the K valley. Note that at the $K^{\prime}$ point, the spin-up and spin-down states are reversed. The arrows schematically indicate the allowed optical transitions leading to A and B excitons in optical spectra. Due to Coulomb-induced electron-hole interaction, the bound excitonic states are located below the conduction band reflecting the corresponding excitonic binding energies. (b) Top view on 
 the hexagonal lattice of $\text{MoS}_\text{2}$ lying in the xy-plane. (c) Side view on the $\text{MoS}_\text{2}$ structure illustrating the Mo layer sandwiched between the two sulfur atom layers. 
}
\label{fig:1}
\end{figure}

We focus on optical transitions between the energetically lowest conduction and the energetically highest valence band. DFT calculations \cite{XiaoPRL2012,ZahidAIP2013,ParhizgarPRB2013,PhysRevBFalko2013,CappellutiPRB2013,LiuPRB2013,OchoaPRB2013} show that in the vicinity of the optically relevant K points, the valence band is mainly formed by the $1/\sqrt{2}\left(d_{x^2+y^2}\pm id_{xy}\right)$ 
orbitals of the molybdenum atoms with a small influence of $1/\sqrt{2}\left(p_{x}\pm ip_{y}\right)$ orbitals of the sulfur atoms. In contrast, the conduction band is dominated by $d_{z^{2}}$ 
orbitals of the Mo atoms with a minor influence of $1/\sqrt{2}\left(p_{x}\pm ip_{y}\right)$ orbitals of the S atoms. 
Here, $+$ and $-$ refer to orbitals forming the K and the K$^\prime$ point, respectively.
For a lattice with $2N$ atoms, we assume the following tight-binding ansatz for the electronic wave function
\begin{align}
 \label{eq:tb}
 \Psi^{\lambda_s \xi}(\mathbf{k},\mathbf{r}) 
 =\dfrac{1}{\sqrt{N}} \sum_{j=\text{Mo,S}}C_{j\mathbf{k}}^{\lambda_s \xi}\sum^N_{\mathbf{R}_{j}} e^{i\mathbf{k}\cdot\mathbf{R}_{j}}\phi_j^{\lambda_s \xi}(\mathbf{r}-\mathbf{R}_{j})
\end{align}
with $\phi_j^{\lambda_s \xi}(\mathbf{r}-\mathbf{R}_{j})$ as the linear combination of the relevant atomic orbitals mentioned above. Here, $\mathbf{R}_{j}$ denotes the coordinates of 
the atoms in the sublattice $j$ built by molybdenum and sulfur atoms, and $\xi$ stands for the K and the K$^\prime$ point, respectively. The coefficients
 $C_j^{\lambda_s \xi}(\mathbf{k})$ determine the weight of the single contributions stemming from different orbital functions. 
 They depend on the two-dimensional momentum $\mathbf{k}$ and the index $\lambda_s$, denoting 
 either the valence ($v_s$) or the conduction bands ($c_s$) with the spin $s=\uparrow$ or $=\downarrow$. 
 The top view on the MoS$_2$ lattice reveals a hexagonal structure that similar to graphene
 consists of two sublattices, cf. Fig. \ref{fig:1} (b). However, while in graphene they are formed by carbon atoms, in MoS$_2$ one sublattice is built by molybdenum and the other by sulfur atoms.
 Therefore, in contrast to graphene the inversion symmetry is broken giving rise to a band gap opening at the K and K$^\prime$ points, cf. Fig. \ref{fig:1}(a). The side view shows that the S atoms build two separate layers with a distance of $\pm 0.15$ nm with respect to the Mo 
 layer, as shown in Fig. \ref{fig:1} (c).\\

To obtain the electronic dispersion relation we solve the Schr\"odinger equation $H_0\Psi^{\lambda_s \xi}=\varepsilon^{\lambda_s}\Psi^{\lambda_s \xi}$ with the Hamilton operator $H_0=H_{\text{kin}}+H_{SO}$. Here, $H_{kin}=\hat{p}^2/(2m_0)$ 
describes the free-particle energy with the momentum operator $\hat{p}$ and the free electron mass $m_0$, while $H_{SO}=U(\mathbf{r})\mathbf{L}\cdot \mathbf{S}$ denotes the spin-orbit coupling (SOC) within the two-center approximation. Here, $\mathbf{L}$ and $\mathbf{S}$ stand for the momentum and the spin operator, respectively, while $U(\mathbf{r})=\frac{1}{2m_0 c^2 r}\frac{d V_r}{d r}$
describes
 the radial dependence of the SOC Hamilton
operator with $V_r$ corresponding to the spherical electrostatic potential and $c$ denoting the speed of light. Since the explicit form of the radial dependence of the molybdenum and sulfur orbitals is unknown, the integration of the radial component will be fixed to the values obtained from experimental data, as discussed below.

Since the orbitals have different symmetries at the K and K$^\prime$ point, we solve the Schr\"odinger equation around both points separately. 
The two sublattice lead to a set of four linear equations. In our nearest-neighbor approach we assume that $\langle\phi_j^{\lambda_s \xi}(\mathbf{r}-\mathbf{R}_{j})|\phi_i^{\lambda_s \xi}(\mathbf{r}-\mathbf{R}_{i})\rangle=\delta_{j,i}$ is a good approximation, i.e. the overlap of orbital functions of neighboring sides is neglected. 
Furthermore, we take into account only the nearest-neighbor hopping integrals $t^{\lambda_s}=\langle\phi_j^{\lambda_s\xi}(\mathbf{r}-\mathbf{R}_{j})|H_{\text{kin}}|\phi_i^{\lambda_s\xi}(\mathbf{r}-\mathbf{R}_{i})\rangle$. Then, we obtain an analytical expression for the electronic bandstructure in the vicinity of the K and K$^\prime$ points reading
\begin{align}
 \epsilon^{\lambda_s}_{\mathbf{k},\xi}=\pm\dfrac{1}{2}\sqrt{\left(\Delta \varepsilon^{\lambda_s}_{\xi}\right)^2+4 |t^{\lambda_s}|^2 f(\mathbf{k})}.
 \label{eq:energy}
\end{align}
with $\xi$ denoting the solution for the K and K$^\prime$ valley.
Furthermore, $\lambda_s= v_\uparrow, v_\downarrow, c_\uparrow, c_\downarrow$ stands for the valence ($v$) and the conduction ($c$) band as well as the spin-up ($\uparrow$) and the spin-down states ($\downarrow$). The $+$ solution is valid for the conduction and the $-$ solution for the valence bands. 

 The momentum dependence of the bandstructure is given by the function
\begin{align}
f(\mathbf{k})=3+2\cos\left(k_y \right)\notag+4\cos\left(3/2k_y\right)\cos\left(\sqrt{3}/2k_x\right) 
\end{align}
stemming from the phase $e^{i\bm k \cdot \bm R_j}$ in the tight-binding wave function in Eq. (\ref{eq:tb}). 
Here $k_x, k_y$ are the Cartesian coordinates of the two-dimensional momentum $\mathbf{k}$ given in units of the lattice vector $a_0=0.318$ nm.\cite{RamasubramaniamPRB2012}
The function contains the trigonal warping effect describing the deviation of the 
 equi-energy contour from a circle around the K and K$^\prime$ points in the Brillouin zone.\cite{ZahidAIP2013,ParhizgarPRB2013,PhysRevBFalko2013,CappellutiPRB2013}
 Since this effect does not have a qualitative influence on excitonic effects, we simplify the electronic bandstructure from Eq. (\ref{eq:energy}) by using a Taylor expansion for small momenta leading to
\begin{align}
\label{eq:BandAppr}
 \epsilon^{\lambda_s}_{\mathbf{k},\xi}\approx
 \pm\left(\dfrac{\Delta \varepsilon^{\lambda_s}_{\xi}}{2}+
 \frac{|t^{\lambda_s}|^2}{\Delta \varepsilon^{\lambda_s}_{\xi}} \mathbf{k}^2\right).
\end{align}
This parabolic bandstructure has already been shown to be a good approximation in the optically relevant energy region around the K and K$^\prime$ points.\cite{XiaoPRL2012,OchoaPRB2013} The tight-binding hoping integrals $t^{\lambda_s}$ determine the curvature of the electronic bandstructure, cf. Eq. (\ref{eq:BandAppr}). The values 
$t^{v_s}=1.25$ eV and $t^{c_s}=1.43$ eV are fixed in such a way that we obtain first-principle values\cite{ZahidAIP2013,ParhizgarPRB2013,PhysRevBFalko2013,CappellutiPRB2013} for the effective mass of the valence band $m_{\text{eff}}^{v}=0.62m_0$ and of the conduction band $m^{c}_{\text{eff}}=0.48 m_0$.
The spin-dependent band gap $\Delta \varepsilon^{\lambda_s}_{\xi}=\varepsilon_{\text{gap}}+\xi\varepsilon_{soc}^{\lambda_s}$ consists of the band gap energy $\varepsilon_{\text{gap}}$ and the spin-orbit splitting $\varepsilon_{soc}^{\lambda_s}$, where $\xi=+, -$ denotes the K and K$^\prime$ point, respectively. As a result, the spin-up (down) electronic state is energetically raised (lowered) by the spin-orbit coupling at the K point and lowered (raised) at the K$^\prime$ point.\cite{HeinzPRL2010,RadisavljevicBNatNano2011,RadisavljevicBNatNano2011}
The broken inversion symmetry in MoS$_2$ gives rise to the spin-independent band gap $\varepsilon_{\text{gap}}=2.41$ eV that is given by the on-site energy difference of the molybdenum and the sulfur atoms.\cite{SanchezPRB2013}
 Both the valence and the conduction band are split due to the efficient spin-orbit coupling, however, the underlying processes are of different order.\cite{OchoaPRB2013}  Consequently, the valence band splitting of $\varepsilon_{soc}^{v_s}=160$ meV is two orders of magnitude larger than the conduction band splitting of $\varepsilon_{soc}^{c_s}=3$ meV.\cite{HeinzPRL2010,Splendiani2010,HeinzNatNano2012,PhysRevBFalko2013} 
Figure \ref{fig:1} illustrates the obtained electronic bandstructure in the exemplary region around the K point. It consists of four parabolic bands stemming from the strong spin-orbit coupling that splits the valence band in two separate spin-up and spin-down bands. 
We find that the spin-orbit coupling also renormalizes the effective masses of spin-up and spin-down valence bands leading to $m_{\text{eff}}^{v_{\downarrow}}=0.66$ and $m_{\text{eff}}^{v_{\uparrow}}=0.575$, cf. Eq.(\ref{eq:BandAppr}). This is in good agreement with the results of Korm\'anyos at al.\cite{PhysRevBFalko2013}

Solving the Schr\"odinger equation, we also obtain the eigenfunctions of electrons in $\text{MoS}_\text{2}$. Assuming that they are normalized, we find for the tight-binding coefficients
\begin{align*}
C_{Mo,\mathbf{k}}^{\lambda_s\xi}=C^{\lambda_s}_{S,\mathbf{k}}g_{\mathbf{k}}^{{\lambda_s}\xi},&&
C_{S,\mathbf{k}}^{\lambda_s \xi}=
\dfrac{\pm1}{\sqrt{1+|g_{\mathbf{k}}^{{\lambda_s}\xi}|^2}},
\label{eq:coef}
\end{align*}
where $g_{\mathbf{k}}^{{\lambda_s}\xi}=t^{\lambda_s} e(\mathbf{k})/(\frac{\Delta\varepsilon^{\lambda_s}_{\xi}}{2}-\varepsilon^{\lambda_s}_{\mathbf{k},\xi})$
and 
$
e(\mathbf{k})=\sum_{j}^3 e^{i\mathbf{k}\cdot \mathbf{b}_j}
$ with $\mathbf{b}_j$ connecting the nearest-neighbor atoms, cf. Fig. 1(b).
Having determined the wave functions, we can now calculate the coupling elements. The carrier-light matrix element $\mathbf{M}^{v_s c_s \xi}({\mathbf k})=\langle\Psi^{v_s \xi}(\mathbf{k},\mathbf{r})|\mathbf p|\Psi^{c_s \xi}(\mathbf{k},\mathbf{r})\rangle$ is given as the expectation value of the momentum operator $\mathbf p =-i\hbar \mathbf{\nabla}$.\cite{malic06b,ErminsBuch}
Exploiting the nearest-neighbor tight-binding wave functions (cf. Eq. (\ref{eq:tb})), we obtain the Cartesian components of $\mathbf{M}^{v_sc_s \xi}({\mathbf k})$ in x(y)-direction 
\begin{align}
\notag
M_{x (y)}^{v_s c_s \xi}(\mathbf{k})&=
 MC_{Mo,\mathbf{k}}^{v\xi *} C_{S,\mathbf{k}}^{c_s\xi}\sum_{l}^{3}b_{l, x (y)} e^{i\mathbf{k}\cdot\mathbf{b}_{l}}\\&
-MC_{S,\mathbf{k}}^{v\xi *} C_{Mo,\mathbf{k}}^{c_s\xi}\sum_{l}^{3}b_{l, x (y)} e^{-i\mathbf{k}\cdot\mathbf{b}_{l}}
\end{align}
with the abbreviation $M=\frac{e\sqrt{3}\hbar}{a_0}\langle\phi_j^{v_s\xi}(\mathbf{r}-\mathbf{R}_{j})|\frac{\partial}{\partial x}| \phi_i^{c_s\xi}(\mathbf{r}-\mathbf{R}_{i})\rangle$ for the nearest-neighbor orbital overlap. 
 While molybdenum orbitals forming the valence and the conduction bands have the d symmetry, the sulfur orbitals are of p-type. 
 Due to the Laporte rule,\cite{LAPORTE25} we expect the main contribution to the optical absorption to stem from transitions between p and d orbitals.
For such transitions, the azimulthal quantum number changes by one, i.e. $\Delta m_l=\pm1$. They can be optically excited by applying circularly polarized light, which carries an azimulthal quantum number of $m_l=\pm1$ depending on the polarization direction. 
In this study, we focus on the qualitative features in the optical spectra of $\text{MoS}_\text{2}$, i.e. we do not aim for the absolute values of the carrier-light coupling. 
Since we are interested on the momentum dependence of the optical matrix element and since we want to keep the number of tight-binding parameters as small as possible, we set all overlap integrals appearing in $M$ to one.

\begin{figure}[t!]
 \begin{center}
\includegraphics[width=0.95\linewidth]{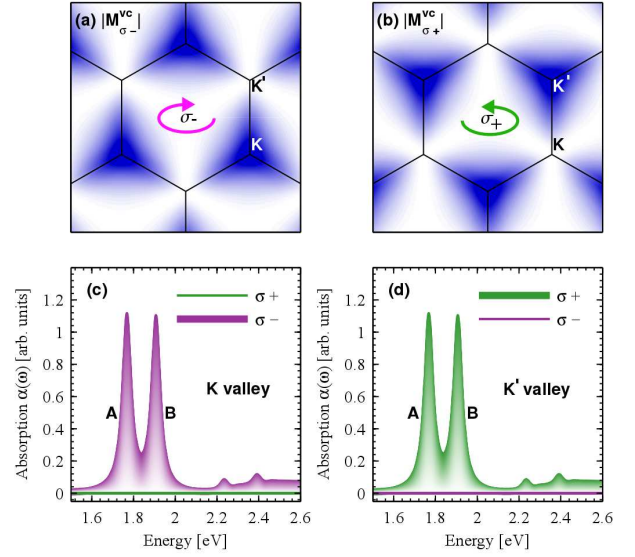}
 \end{center}
 \caption{The optical matrix element projected in the direction of (a) right- ($\sigma-$) and (b) left-handed ($\sigma+$) circularly polarized light. The corresponding absorption spectra in the spectral region of the (c) K and (d) K$^\prime$ valley after optical excitation with $\sigma-$ and $\sigma+$ light, respectively. The figure illustrates a pronounced valley-selective polarization, i.e. the excitation with right-(left-)handed circularly polarized light only leads to an absorption at the K (K$^\prime$) point. 
 \label{fig:2}
}\end{figure}

The excitation pulse is characterized by the vector potential $\mathbf A = \mathbf A_0 \exp(\frac{t^2}{2\sigma_t^2})(\cos(\omega t)\mathbf e_x+\sin(\omega t)\mathbf e_y)$ with the amplitude $A_0$ determining the excitation strength and $\sigma_t$ denoting the pulse width.
 Projecting the optical matrix element in the polarization direction, we can express the carrier-light coupling by a linear combination of the Cartesian components $M_{x}^{v_sc_s \xi}(\mathbf{k})$ and $M_{y}^{v_sc_s \xi}(\mathbf{k})$. For right- ($\sigma_-$) and left-handed ($\sigma_+$) circularly 
polarized light, we find $M_{\sigma_\pm}^{v_sc_s \xi}(\mathbf{k})=M_{x}^{v_sc_s \xi}(\mathbf{k})\pm iM_{y}^{v_sc_s \xi}(\mathbf{k})$.\cite{CaoTing2012}
 Figures \ref{fig:2} (a) and (b) illustrate the optical matrix elements $M_{\sigma_+}^{v_sc_s \xi}(\mathbf{k})$ and $M_{\sigma_-}^{v_sc_s \xi}(\mathbf{k})$, respectively. We can clearly observe that the carrier-light coupling strongly
differs for different polarization of the light: For right-handed circularly polarized light $\sigma_-$, it exhibits maxima at the 
 K points, while it vanishes at the K$^\prime$ points, cf. Fig. \ref{fig:2}(a). The behavior is inverse for left-handed circularly polarized light $\sigma_+$, as shown in Fig. \ref{fig:2}(b).
 Furthermore, the matrix element shows similar to the dispersion relation a strong trigonal warping effect reflecting the three-fold symmetry of the nearest-neighbors in the real space lattice, cf. Fig. \ref{fig:1}(b). The appearing triangles in the optical matrix element show a different orientation depending on the polarization of light.

Figures \ref{fig:2} (c) and (d) illustrate the absorption spectrum of $\text{MoS}_\text{2}$ in the spectral region of the K and K$^\prime$ valley, respectively, after optical excitation with right- ($\sigma_-$) and left-handed ($\sigma_+$) circularly polarized light. 
The carriers occupying the K valley only couple to the $\sigma_-$ light leading to pronounced peaks in the absorption spectrum. The excitation with $\sigma_+$ light does not lead to any absorption due to the vanishing optical matrix element, cf. Fig. \ref{fig:2}(a). In contrast, at the K$^\prime$ valley, the behavior is reverse
and we only observe pronounced peaks after the excitation with $\sigma_+$ light. This valley-selective polarization has also been observed in experiments and enables the full control of the valley and spin occupation by optical excitation with circularly polarized light suggesting the application of $\text{MoS}_\text{2}$ in valley-tronics.\cite{CaoTing2012,HeinzNatNano2012,MakKinFai2013,Hualing2012,Sanfeng2013,SallenPRB2012}\\

The absorption spectra in Fig. \ref{fig:2} (c) and (d) are characterized by strongly pronounced excitonic Lorentzian-shaped resonances reflecting the efficient Coulomb interaction in $\text{MoS}_\text{2}$. Similar to graphene\cite{malic11} or carbon nanotubes,\cite{malic10,malic10b} the Coulomb interaction is known to be particularly important in monolayer TMDs due to their low-dimensionality and the relatively weak screening in such one-atom thick materials. Therefore, we extend the Hamilton operator by the Coulomb interaction to account for the Coulomb-induced features in the absorption spectrum yielding in second quantization \cite{Kochbuch}
\begin{align*}
H_{c}=&\frac{1}{2}\sum_{\mathbf{k},\mathbf{k'},\mathbf q}\sum_{\lambda_s\lambda_s'}
a^{\dagger}_{\mathbf{k},\lambda_s}a^{\dagger}_{\mathbf{k'},\mathbf q,\lambda_s'}
a_{\mathbf{k'}+\mathbf{q},\lambda_s'}a_{\mathbf{k}-\mathbf{q},\lambda_s}
V_{\mathbf{k},\mathbf{k'},\mathbf{q}}^{\lambda_s\lambda_s'\xi}
\end{align*}
with the Coulomb matrix element
\begin{align*}
V_{\mathbf{k},\mathbf{k'},\mathbf{q}}^{\lambda_s\lambda_s'\xi}=\varGamma^{\lambda_s\lambda_s'\xi}_{\mathbf{k},\mathbf{k'},\mathbf{q}}V_{\mathbf{q}}^{2D}\zeta(\mathbf{q}), 
\end{align*}
the annihilation and creation operators $a^{\dagger}_{\mathbf{k},\lambda_s}
$ and $a_{\mathbf{k},\lambda_s}$, the momenta $\mathbf k,\mathbf k'$ and the band indices $\lambda_s,\lambda_s'$ of the involved electronic states.
The Coulomb matrix element is determined by the tight-binding coefficients
\begin{align*}
\varGamma^{\lambda\lambda'\xi}_{\mathbf{k},\mathbf{k'},\mathbf{q}}=\sum_{j,f=S,Mo}C_{f,\mathbf{k}}^{\lambda_s\xi *}C_{j,\mathbf{k'}}^{\lambda'\xi *}
C_{j,\mathbf{k'}+\mathbf{q}}^{\lambda'\xi}C_{f,\mathbf{k}-\mathbf{q}}^{\lambda_s\xi}
\end{align*}
and the Fourier-transformed two-dimensional Coulomb potential 
$V_{\mathbf{q}}^{2D}=\frac{e^2}{2\epsilon_0\epsilon_rL^2}\frac{1}{|\mathbf{q}|}$ including the sample size $L^2$, the vacuum permittivity $\epsilon_0$ 
 and the dielectric constant $\epsilon_r=3.4$ describing the screening within the $\text{MoS}_\text{2}$ monolayer.\cite{CheiwchanchamnangijPRB2012} 

To account for the thickness of $\text{MoS}_\text{2}$, we consider
a confinement function $\zeta(\mathbf{q})$ corresponding to the integral over the envelope functions perpendicular to the lattice. The effective thickness $L_\bot\approx 0.66$ nm of the $\text{MoS}_\text{2}$ monolayer is approximated by taking into account the distance between the two sulfur atoms in the direction perpendicular to the layer ($0.3$ nm) and the Van der Waals diameter of sulfur ($0.36$ nm).\cite{Hongyan2013}

Now, we have all ingredients including the electronic bandstructure, the optical coupling element, and the Coulomb matrix element to calculate the excitonic absorption spectrum of $\text{MoS}_\text{2}$. To obtain the absorption coefficient $\alpha(\omega)$, we need to know the temporal evolution of the microscopic polarization $p^{c_sv_s}_{\mathbf{k}}=\langle a^{\dagger}_{\mathbf{k},c_s}a_{\mathbf{k},v_s}\rangle$ 
which is a measurement for the optical transitions of electrons in the state $\mathbf k$ between the valence ($v_s$) and the conduction ($c_s$) band.\cite{malic11,ErminsBuch}
Exploiting the Heisenberg equation of motion,\cite{Kochbuch} we 
derive the semiconductor Bloch equation for the microscopic polarization yielding
\begin{align}
\label{eq:x}
 i\hbar \dot{p}_{\mathbf{k}, \xi}^{c_sv_s}(t)=
 \tilde{\epsilon}^{s}_{\mathbf{k},\xi} p_{\mathbf{k},\xi}^{c_sv_s}(t)-\left(f^{c_s}_{\mathbf{k}\xi}-f^{v_s}_{\mathbf{k}\xi}\right)\tilde{\Omega}_{\mathbf{k}}(t).
\end{align}
Here, we used the Cluster expansion to truncate the many-particle hierarchy problem on the Hartree Fock level.\cite{Kochbuch,PQE} 
The Coulomb interaction leads to the renormalization of the band gap energy resulting in
$\tilde{\epsilon}^{s}_{\mathbf{k},\xi}=
\epsilon^{c_s}_{\mathbf{k},\xi}-\epsilon^{v_s}_{\mathbf{k},\xi}-\sum_{\mathbf{k'}}\left(f^{c_s}_{\mathbf{k}\xi}V_{\mathbf{k},\mathbf{k'},\mathbf{q}}^{c_sc_s}-f^{v_s}_{\mathbf{k}\xi}V_{\mathbf{k},\mathbf{k'},\mathbf{q}}^{v_sv_s}\right)$ and to the renormalization of the Rabi frequency resulting in
$\Omega_\mathbf{k}=\mathbf{M}_{\pm\mathbf k }^{v_s c_s\xi}\cdot\mathbf{A}(t)+\sum_{\mathbf{k'}}V_{\mathbf{k},\mathbf{k'},\mathbf{q}}^{exc}p_{\mathbf{k'}, \xi}^{c_sv_s\xi}(t)$. Here, we introduced the abbreviation 
$V_{\mathbf{k},\mathbf{k'},\mathbf{q}}^{exc}=V_{\mathbf{k},\mathbf{k'},\mathbf{q}}^{\lambda_s\lambda_s'\xi}$ for $\lambda_s\neq\lambda_s'$ expressing the electron-hole contribution of the Coulomb interaction.
Focusing on linear optics, where the optical perturbation is weak, we can assume for an undoped system in equilibrium $f^{c_s}_{\mathbf{k}\xi}=0$ and $f^{v_s}_{\mathbf{k}\xi}=1$ neglecting thermal occupations.
Since the band gap energy has been fixed according to first-principle calculations including the GW approximation,\cite{SanchezPRB2013} the Coulomb-induced energy renormalization is already taken into account. 

To get insights into the intrinsic properties of the system, we fist investigate the homogeneous solution of Eq. (\ref{eq:x}), which defines the eigenvalue problem
\begin{align}
\label{eq:Wannier}
 \tilde{\epsilon}^{s}_{\mathbf{k},\xi} \theta_{\nu\xi}^s(\mathbf{k})-\sum_{\mathbf{k'}}V_{\mathbf{k},\mathbf{k'},\mathbf{q}}^{exc}\theta_{\nu\xi}^{s}(\mathbf{k}^\prime)=E_{\nu\xi}^{s}\theta_{\nu\xi}^s(\mathbf{k})
\end{align}
corresponding to the well-known Wannier equation.\cite{Kochbuch,PQE} The eigenvalues $E_{\nu\xi}^{s}$ present solutions of the excitonic problem giving access to the spectral position as well as the binding energies of the s-like excitonic states. The excitonic wave functions $\theta_{\nu\xi}^{s}(\mathbf{k})$ determine the oscillator strength of the excitonic transitions appearing in the absorption spectrum. 
 \begin{figure}[t!]
 \begin{center}
\includegraphics[width=0.95\linewidth]{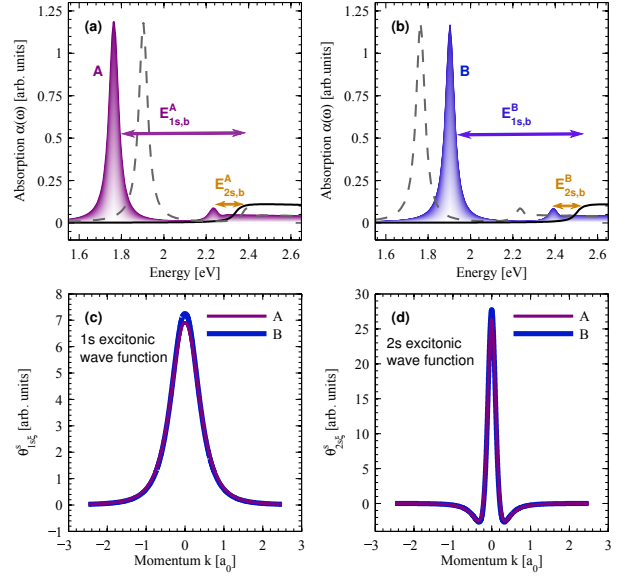}
 \end{center}
 \caption{Absorption spectrum of the free-standing $\text{MoS}_\text{2}$ focusing on the well-pronounced (a) A and (b) B exciton arising from the transition from the two
energetically highest spin-split valence bands to the energetically lowest
conduction band, respectively. The corresponding excitonic binding energies 
$E^{A/B}_{1s,b}$ can be directly read off from the difference to the free-particle transitions, cf. the black lines. Note that we have doubled the free-particle absorption intensity for better visibility throughout the paper. Higher low-intensity excitonic transitions with binding energies $E^{A/B}_{2s,b}$ can also be observed. The corresponding eigenfunctions $\theta_{\nu\xi}^{s}(\mathbf{k})$ of 1s and 2s excitons are shown in (c) and (d), respectively. The determine the oscillator strength of the excitonic transitions. 
\label{fig:3}
}\end{figure} 
The absorption coefficient $\alpha(\omega)$ is proportional to the imaginary part of the susceptibility $\chi(\omega)$, which can be expressed via the macroscopic current density $j(\omega)$.\cite{ErminsBuch} The latter is directly determined by the microscopic polarization $p_{\mathbf{k}\xi}^{c_sv_s}(t)$.
With the solution of Eq. (\ref{eq:Wannier}), we can express the microscopic polarization by transforming 
Eq.(\ref{eq:x}) using the relations $p_{\mathbf{k}\xi}^{c_sv_s}(t)=\sum_{\nu}p_{\nu\xi}^{c_sv_s}(t) \theta_{\nu\xi}^{s}(\mathbf k)$ and $p_{\nu\xi}^{c_sv_s}(t)=\sum_{\nu}p_{\mathbf{k}\xi}^{c_sv_s}(t) \theta_{\nu\xi}^{s*}(\mathbf k)$.\cite{PQE}
The new quantity $p_{\nu\xi}^{c_sv_s}(t)$ depends on the excitonic eigenvalues and can be expressed analytically in the frequency domain yielding
\begin{align}
 p_{\nu\xi}^{c_sv_s}(\omega)=\frac{\sum_{\mathbf{k}}M_{\pm\mathbf k }^{v_s c_s\xi}A(\omega)\theta_{\nu\xi}^{s}(\mathbf k)}{E_{\nu\xi}^s-\hbar\omega-i\gamma}.
\end{align}
Finally, we obtain for the absorption coefficient the analytical expression \cite{Kochbuch,PQE}
\begin{align}
\label{eq:Elliot}
\alpha(\omega)=\frac{1}{\epsilon_0\epsilon_r\omega}\Im\left[{\sum_{\nu\xi,s}\dfrac{\Theta_{\nu\xi}^s}{E_{\nu\xi}^s-\hbar\omega-i\gamma}}\right].
\end{align}
 This equation corresponds to the Elliot formula, which describes the macroscopic answer of the system to an external optical perturbation.\cite{Kochbuch,PQE} Note that we have introduced a phenomenological dephasing rate $\gamma=25$ meV to account for higher correlation terms neglected on the Hartree Fock level. This rate determines the width of transition peaks appearing in the absorption spectrum, however, it does not have any influence on their position or the excitonic binding energy.
We find that the oscillator strength of the peaks in the absorption spectrum is determined by the square of the optical matrix element $M_{\pm}^{v_sc_s}(\lambda,\phi)$ and the sum over excitonic wave functions $\Theta_{\nu\xi}^{s}=\sum_{\mathbf{k}}\theta_{\nu\xi}^{s}(\mathbf{k},\sigma)M_{\pm}^{v_sc_s\xi}(\mathbf{k})\sum_{\mathbf{k'}}[\theta_{\nu\xi}^{s}(\mathbf{k'})M_{\pm}^{v_sc_s\xi}(\mathbf{k'})]^*$
The eigenvalues $E_{\nu\xi}^s$ of Eq.(\ref{eq:Wannier}) appearing in the denominator of the Elliot formula determine the position of the excitonic peaks as well as their binding energy.

The absorption spectrum of $\text{MoS}_\text{2}$ features pronounced peaks clearly arising from excitonic transitions, as free-particle band-to-band
transitions in a two-dimensional material give steps in absorption, cf. Figs. \ref{fig:3}(a) and (b).
The appearing two peaks stem from transitions between the two
energetically highest spin-split valence bands to the energetically lowest
conduction band, cf. Fig. \ref{fig:1}(a).
The energetically lower (higher) transition 
is denoted as the A (B) exciton in literature.\cite{HeinzPRL2010}
In the case of free-standing $\text{MoS}_\text{2}$, i.e. without considering a substrate-induced dielectric background screening of the Coulomb potential, the A exciton is located at 1.76 eV and the
B exciton at 1.9 eV.
Compared to the experimental data \cite{MakKinFai2013}, the peak position is red-shifted by approximately 150 meV. This can be traced back to the impact of the substrate, as discussed below.
We also calculate the Coulomb-renormalized band-to-band transitions in the absorption spectrum to be able to determine the excitonic binding energies $E^{A}_{1s,b}=570$ meV and $E^{B}_{1s,b}=590$ meV, cf. the arrows in Figs. \ref{fig:3}(a) and (b). The difference of 20 meV can be traced back to the different effective masses of the spin-split valence bands.

Besides the two main A and B peaks, we also observe further higher excitonic resonances with a much smaller intensity. In analogy to the Rydberg series in the hydrogen atom, each exciton transition splits into a series of optically active exciton states. In the absorption spectrum of $\text{MoS}_\text{2}$, we observe the 2s excitonic resonances that are located at 0.47 eV and 0.49 eV above the A and B excitons (corresponding to the 1s transitions), respectively. They show a weak intensity that is by one magnitude smaller than the corresponding 1s transitions. Their excitonic binding energy $E^{A/B}_{2s,b}$ is in the range of 100 meV.
 To investigate the relative oscillator strength of the observed peaks, we plot the excitonic eigenfunctions $\theta_{\nu\xi}^{s}(\mathbf{k})$ found as solution of Eq. (\ref{eq:Wannier}). Figure \ref{fig:3}(c) reveals that the eigenfunction of the B exciton is slightly higher. We can trace this behavior back to the difference in the 
effective mass $m^*_{\lambda_s}$ of the involved electronic bands $\lambda_s$. Our calculations show that the oscillator strength is enhanced for increasing $m^*_{\lambda_s}$. Due to the spin-orbit coupling, the effective mass of the energetically higher valence band is larger giving rise to a larger oscillator strength of the B exciton.
 However, this effect is almost completely canceled due to the $1/\omega$-dependence of the absorption coefficient (cf. Eq. (\ref{eq:Elliot})), which suppresses energetically higher transitions. As a result, the absorption spectrum shows that both peaks have nearly the same oscillator strength.

 \begin{figure}[t!]
 \begin{center}
\includegraphics[width=0.95\linewidth]{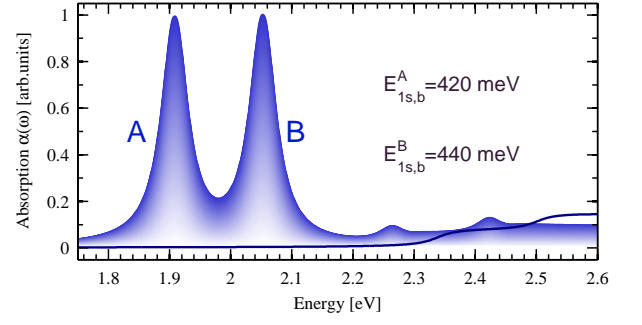}
 \end{center}
 \caption{Excitonic absorption spectrum of $\text{MoS}_\text{2}$ on a silicon substrate. Compared to free-standing molybdenum disulfide shown in Fig. \ref{fig:3}, the two excitonic resonances are blue shifted by approximately $150$ meV due to the substrate-induced screening of the Coulomb interaction. The peak positions at 1.91 eV 
and 2.05 eV are in good agreement with the experiment.\cite{MakKinFai2013} The corresponding excitonic binding energies are $E^A_{1s,b}=420$ and $E^B_{1s,b}=440$ meV.}
 \label{fig:4}
\end{figure} 
 
To compare our results with the recent experimental data,\cite{MakKinFai2013} we study the absorption spectrum of $\text{MoS}_\text{2}$ on a silicon substrate characterized by a dielectric background constant
of $\epsilon_r=4.1$ \. The latter gives rise to an efficient screening of the Coulomb potential affecting the position and the binding energy of excitonic transitions. 
The corresponding absorption spectrum is shown in 
Fig.\ref{fig:4}. We find that the excitonic resonances are blue shifted by approximately $150$ meV. As a result, the A exciton is located at 1.91 eV and the B exciton at 2.05 eV, which is in very good agreement with the experimental observation.\cite{MakKinFai2013} We could not reproduce the measured relative oscillator strength of the A and B excitons. While in the experiment, the A exciton is higher in intensity, our theoretical spectra show nearly the same oscillator strength for both excitons. This might be due to the higher-order effects beyond the considered Hartree-Fock approximation and will be studied in future work. 
Figure \ref{fig:4}, furthermore, shows that the substrate-induced screening also influences the excitonic binding energies. Compared to free-standing $\text{MoS}_\text{2}$, they are reduced to $E^A_{1s,b}=420$ and $E^B_{1s,b}=440$ meV. Our results are clearly smaller than the predicted values of 1 eV in Ref. \onlinecite{RamasubramaniamPRB2012} and are rather in agreement with the more recent well-converged first-principle study by A. Molina-Sanchez et al.\cite{SanchezPRB2013}\\ 
In conclusion, we have presented an analytical description of the excitonic absorption spectrum of the $\text{MoS}_\text{2}$ monolayer. Our approach is based on the density matrix formalism allowing a consistent treatment of the carrier-light and carrier-carrier interaction on microscopic footing. 
We investigate the formation of bound electron-hole pairs and their influence on the absorption spectrum of $\text{MoS}_\text{2}$. In agreement with experimental data, our calculations show the possibility of valley-selective polarization as well as the appearance of strongly pronounced A and B excitons with binding energies in the range of few hundreds of meV. Furthermore, we predict the occurrence of still unobserved higher excitonic transitions characterized by much lower intensities. Moreover, we investigate the impact of the excitonic eigenfunctions on the relative oscillator strength of the excitonic peaks as well as the influence of substrate-induced screening on the excitonic binding energies. Our approach can be applied to the optical properties of other transition metal dichalcogenides as well as extended to investigations of the non-equilibrium carrier dynamics beyond the Hartree-Fock level.\cite{winzer10, winzerPRB2013}

We acknowledge the financial support from the Einstein Foundation Berlin. We thank Andreas Knorr and Florian Wendler for inspiring and fruitful
discussions.

\end{document}